\documentclass[]{interact}

\usepackage[T1]{fontenc}
\usepackage{lmodern}
\usepackage{graphicx}
\usepackage{booktabs}
\usepackage{amsmath,amssymb,amsfonts}
\usepackage{float}
\usepackage{placeins}
\usepackage{hyperref}

\usepackage[longnamesfirst,sort]{natbib}
\bibpunct[, ]{(}{)}{;}{a}{,}{,}

\begin{document}

\title{Detective scaffolding for within-session reasoning development: a three-phase framework evaluated in polymer engineering and pre-university outreach}

\author{
\name{Haolin Feng\textsuperscript{a}, Holly Barrett\textsuperscript{a}, Xinru Deng\textsuperscript{a}, Dimitrios G.\ Papageorgiou\textsuperscript{a} and Yiwei Sun\textsuperscript{a}\thanks{CONTACT Yiwei Sun. Email: yiwei.sun@qmul.ac.uk}}
\affil{\textsuperscript{a}School of Engineering and Materials Science, Queen Mary University of London, London, United Kingdom}
}

\maketitle

\begin{abstract}
This paper presents a detective scaffolding framework---a three-phase
instructional sequence (Hypothesis Activation $\rightarrow$ Evidence
Structuring $\rightarrow$ Causal Integration) in which engineering
students investigate a realistic industrial defect scenario using
staged in-class polls as designed evidence probes. Unlike conventional
uses of student response systems for engagement, the framework
positions each poll as an Evidence-Centred Design instrument targeting
a specific reasoning capability. In the primary implementation,
80 Year~3 polymer engineering students progressed from
prior-knowledge-driven misconception (71\% attributing defects to
temperature) to complete root-cause convergence (100\% identifying
humidity; Fisher's exact test, $p < .001$) across four sequenced
prompts within a single 90-minute lecture slot. A dual-accuracy
analysis revealed that at one intermediate stage, textbook-correct and
analytically valid responses diverged, illustrating why conventional
scoring can misrepresent reasoning quality. In a transferability
study, 26 Year~12 students with no engineering background achieved
identical root-cause identification rates across two adapted scenarios,
with significant gains in data-analysis confidence and AI explanation
ability. The results suggest that the pedagogical structure, rather
than disciplinary content, drives the convergence effect, implying
portability across disciplines and educational levels.
\end{abstract}


\section{Introduction}\label{sec:intro}

Active learning is among the most robustly supported interventions in
science education.  A meta-analysis of 225 studies found that active
learning reduced failure rates by 33\% and increased examination
performance by half a standard deviation across STEM disciplines
\citep{Freeman2014}, and a subsequent synthesis showed that these
benefits disproportionately narrow achievement gaps for
underrepresented students \citep{Theobald2020}.  Within engineering
specifically, active, collaborative, and cooperative methods have
long been recognised as more effective than lecture alone
\citep{Prince2004}.  The question facing engineering educators is
therefore no longer \emph{whether} active learning works, but
\emph{how to design activities that make student reasoning visible
during the learning process itself, and whether those designs
transfer across contexts}.

Most evaluations of active learning rely on endpoint measures: final
examination scores, course grades, or post-intervention surveys
\citep{Freeman2014}.  These instruments confirm that learning
occurred but cannot reveal the trajectory by which it occurred.
Within a single session, student reasoning may shift through
qualitatively distinct phases---from prior-knowledge-driven
misconception, through evidence evaluation, to causal synthesis---yet
this progression is typically invisible to both instructors and
researchers.  The formative assessment literature has increasingly
argued that what matters is not generic participation but the
\emph{disciplinary substance} of the thinking being elicited
\citep{Coffey2011}; sequential analysis of classroom interactions has
been proposed as one way to capture these dynamics
\citep{Furtak2017}; and Evidence-Centred Design (ECD) provides a
principled framework for designing tasks that generate interpretable
evidence about student competence \citep{Mislevy2004, Kubsch2022}.
Yet in practice, student response systems are predominantly studied
as engagement and feedback tools \citep{Caldwell2007, Pai2025,
Ranjbaran2023} rather than as structured evidentiary sequences that
trace reasoning progression within a learning activity.

A second gap concerns the pedagogical framing of investigation.
Gamified and narrative-driven approaches have gained traction in
science education, with escape rooms attracting systematic review
attention \citep{Veldkamp2020} and forensic case studies showing
engagement benefits in chemistry \citep{Dinan2007}.  In polymer
education specifically, puzzle-based formats have been used to teach
structure--property relationships \citep{Gilbert2020}, and
problem-based approaches have addressed characterisation techniques
\citep{HeZou2025} and sustainability \citep{Knight2025}.  However,
these interventions tend to share several constraints: they operate at
small scale (typically 20--30 students in dedicated settings), use
purpose-designed puzzles with artificial causal structures rather than industrially informed scenarios, and
report endpoint outcomes rather than within-session progression.
Meanwhile, the peer instruction tradition has demonstrated that
polling can support large-cohort conceptual learning
\citep{Crouch2001, Smith2009}, but peer instruction is built around
discrete concept questions rather than sustained multi-stage
investigation.  What is missing is a design that combines the
motivational and epistemic affordances of narrative investigation with
the scalability of polling-based pedagogy and the evidential rigour
of formative assessment theory.

A third gap concerns transferability.  Most active learning studies
report single-context implementations; few test whether the same
pedagogical framework produces comparable outcomes across populations
differing in age, prior knowledge, and educational level.  Yet the
question of whether learning gains derive from a transferable
pedagogical architecture or from topic-specific content is central to
the generalisability of any instructional design.

This paper addresses all three gaps through the design and evaluation
of a \emph{detective scaffolding framework}---a three-phase
instructional sequence (Hypothesis Activation $\rightarrow$ Evidence
Structuring $\rightarrow$ Causal Integration) in which students
investigate a realistic industrial defect scenario using staged
in-class polls as designed evidence probes.  We report findings from
two deployments.  In the primary implementation, 80 Year~3 polymer
engineering students worked in informal teams through four sequenced
Mentimeter prompts while interrogating an injection-moulding dataset
containing over 50~process variables, with a hidden root cause
(ambient humidity) embedded among salient but non-causal factors.  In
a transferability study, 26 Year~12 students with no engineering
background completed an adapted version of the framework in a
four-hour outreach workshop, with an additional AI-scaffolded
investigation sequence.

The study makes four contributions.  First, it provides stage-wise evidence of within-session reasoning development---from initial misconception (71\% selecting temperature) to unanimous selection of humidity/moisture as a key contributing factor among final responding teams---captured through polls functioning as
ECD-informed evidence probes rather than engagement tools.  Second, it
provides a worked example of the dual-accuracy principle: at one
stage, textbook-correct and analytically valid responses diverge,
illustrating why conventional scoring can misrepresent the quality of
student reasoning.  Third, it shows that the detective scaffolding
framework operates at large-cohort scale within a standard lecture
slot (80~students, 90~minutes), addressing a practical constraint
that limits many PBL and escape-room designs \citep{Savery2006,
Kolmos2009}.  Fourth, it provides preliminary evidence of cross-context portability:
pre-university students with no domain expertise achieved identical
root-cause identification rates, with significant gains in data
analysis confidence and AI explanation ability, suggesting that the pedagogical structure may contribute more to the convergence effect than the specific disciplinary content.

\section{Theoretical framework}\label{sec:framework}

The detective scaffolding framework draws on four complementary
theoretical traditions to justify both its structure and its
assessment logic.

\paragraph{Staged cognitive engagement.}
The ICAP framework predicts that learning outcomes improve as students
move from passive reception through active responding and constructive
generation to interactive co-elaboration \citep{Chi2014}.  The
three-phase design operationalises this hierarchy.  In the first phase
(\emph{Hypothesis Activation}), students respond to an initial poll
by selecting from given causal options, drawing on prior knowledge
without yet examining evidence---an \emph{Active}-level task.  In the
second phase (\emph{Evidence Structuring}), they interpret temporal
and structural patterns in the dataset and generate descriptions that
go beyond the information directly presented, which constitutes
\emph{Constructive} engagement.  In the third phase (\emph{Causal
Integration}), students synthesise evidence across preceding stages
to identify the root cause, ideally through within-team discussion
that meets the conditions for \emph{Interactive} engagement.  The
staged design is thus not merely a sequence of questions but a
deliberate escalation through ICAP levels.

\paragraph{Productive failure.}
Kapur's productive failure research demonstrates that students who
first struggle with ill-structured problems before receiving
consolidation show deeper understanding than those who receive direct
instruction from the outset \citep{Kapur2008}.  The framework
exploits this mechanism by design: Phase~1 deliberately invites an
initial misconception (temperature-driven defects) that most students
will select.  This is not a flaw but a pedagogical feature.  The
subsequent phases provide the structured evidence that allows students
to revise their initial hypothesis, mirroring the
generate-then-consolidate sequence that productive failure predicts
should be maximally effective.

\paragraph{Anchored instruction.}
The detective framing provides what the Cognition and Technology Group
at Vanderbilt termed an ``anchor''---a realistic, narrative-rich
problem context that situates reasoning in a complex information
environment \citep{CTGV1993}.  Unlike abstract textbook problems, the
injection-moulding scenario presents over 50~process variables, shift
patterns, and environmental factors, requiring students to
distinguish signal from noise.  The narrative framing (``you are
investigating a factory defect'') serves two functions: it motivates
sustained engagement through a sense of investigative agency, and it
legitimises incorrect intermediate hypotheses as ``leads'' rather
than errors, thereby supporting the productive failure mechanism
described above.

\paragraph{Evidence-Centred Design.}
Mislevy's ECD framework argues that credible assessment requires
explicit alignment between claims about competence, the evidence
needed to support those claims, and the tasks that elicit that
evidence \citep{Mislevy2004}.  Recent extensions connect ECD to
learning progression analytics, arguing that learning environments
should generate interpretable evidence about how understanding
develops \citep{Kubsch2022}.  We apply this logic to in-class
polling: each Mentimeter prompt is designed not as a participation
device but as an evidence probe targeting a specific reasoning
capability (hypothesis generation, temporal analysis, structural
classification, and causal synthesis).  This framing moves polling
beyond engagement measurement toward a structured evidentiary chain
in which each stage's responses reveal whether the targeted reasoning
transition has occurred.

Table~\ref{tab:ecd} makes this logic concrete by mapping each poll
to its target reasoning capability, the claim it tests, and the
evidence that would indicate success.
 
\begin{table}[!htbp]
\centering
\caption{Evidence-Centred Design mapping for the four staged polls.}
\label{tab:ecd}
\small
\begin{tabular}{@{}p{1.1cm}p{2.2cm}p{3.0cm}p{3.0cm}p{3.0cm}@{}}
\toprule
Poll & Phase & Target reasoning & Claim about thinking & Success criterion \\
\midrule
Q1
  & Hypothesis Activation
  & Prior-knowledge hypothesis generation
  & Students default to salient processing variables before examining data
  & High concentration on one dominant option (temperature) \\[6pt]
Q2
  & Evidence Structuring
  & Temporal pattern recognition
  & Students shift from ``what causes defects'' to ``when do defects peak''
  & ${>}50\%$ identify correct time window \\[6pt]
Q3
  & Evidence Structuring
  & Structural classification
  & Students model signal form rather than isolated timing
  & Shift from temporal labels toward step/periodic descriptions \\[6pt]
Q4
  & Causal Integration
  & Root-cause synthesis
  & Students integrate temporal and structural evidence into a causal account
  & ${>}80\%$ select humidity/moisture as a contributing factor \\
\bottomrule
\end{tabular}
\end{table}

\medskip

\begin{figure}[!htbp]
\centering
\includegraphics[width=0.75\linewidth]{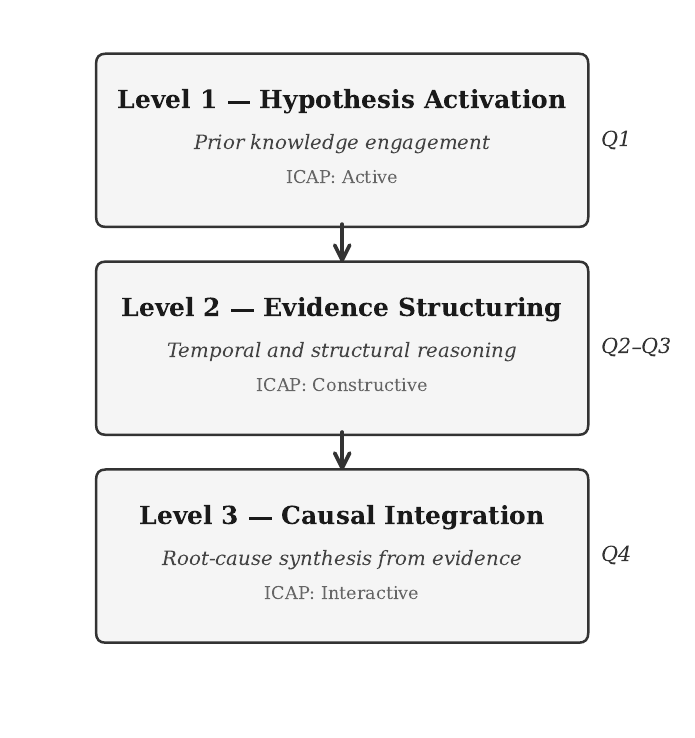}
\caption{Three-phase detective scaffolding framework.
Phase~1 (Hypothesis Activation, Q1) elicits prior-knowledge-driven
causal attributions.
Phase~2 (Evidence Structuring, Q2--Q3) guides students through
temporal and structural analysis of the dataset.
Phase~3 (Causal Integration, Q4) requires synthesis of accumulated
evidence into a root-cause identification.
Arrows indicate the designed reasoning progression; poll stages are
annotated at right.}
\label{fig:framework}
\end{figure}

Figure~\ref{fig:framework} illustrates the resulting three-phase
structure and its mapping to the staged polls.  The framework is
intentionally minimal---three phases, four polls---to preserve
feasibility within a single lecture slot for large cohorts.

\FloatBarrier

\section{Activity design and methods}\label{sec:methods}

This study follows an empirically grounded design-and-evaluation
approach with mixed methods: the detective scaffolding framework was
developed from theory (Section~\ref{sec:framework}), implemented in
two contexts, and evaluated using quantitative polling data,
qualitative participant reflection, and cross-context comparison.

\subsection{Undergraduate implementation}\label{subsec:ug_methods}

\subsubsection{Context and participants}

The Polymer Detective activity was implemented in the Physical
Properties of Polymers module (QXU5032, Year~3) at a joint UK--China
undergraduate programme.  The module covers polymer processing,
structure--property relationships, and failure analysis.  Eighty
students participated during a scheduled 90-minute lecture slot.
Students were not pre-assigned to formal teams; instead, they were
encouraged to discuss with their neighbours and form casual
investigative groups of three to four members, reflecting the
informal collaborative structures typical of large-lecture active
learning \citep{Smith2009}.

\subsubsection{Materials}

Students received access to a purpose-built injection-moulding dataset
modelled on the authors' industrial experience in polymer processing,
containing over 50~process and environmental variables, including
barrel temperature, injection pressure, mould temperature, ambient
humidity, operator shift patterns, and raw material batch codes.  The
dataset featured realistic noise and intercorrelation.  The root cause of the defect
increase---ambient humidity spikes coinciding with shift
changes---was embedded among salient but non-causal variables
(particularly temperature, which students' prior coursework had emphasised as the dominant processing parameter). Barrel and mould temperatures in the dataset were held within normal operating tolerances throughout the production period, such that temperature-driven hydrolysis was not a contributing factor; the defect increase was attributable solely to ambient humidity ingress prior to drying. The scenario represented a facility without pre-drying protocols, meaning pellet drying did not appear as a corrective variable in the dataset. This design
ensured that the most cognitively available initial hypothesis
(temperature) would be incorrect, activating the productive failure
mechanism described in Section~\ref{sec:framework}.

Three versions of a supporting Python analysis script were provided
at different levels of scaffolding (fully guided, partially guided,
and open-ended) to accommodate heterogeneous programming confidence
within the cohort.  Students were free to choose any version or to
explore the dataset using their own methods.

\subsubsection{Procedure}

The session followed the three-phase framework
(Fig.~\ref{fig:framework}).

\emph{Phase~1: Hypothesis Activation} (Q1).  Before examining the
dataset in detail, students were asked via Mentimeter: ``What do you
think is causing the increase in defects?''  Five response options
were provided (material degradation, temperature fluctuations,
moisture in the polymer, operator error, equipment wear).  This poll
captured prior-knowledge-driven causal attributions before any
evidence had been evaluated.

\emph{Phase~2: Evidence Structuring} (Q2--Q3).  Students then worked
through the dataset in their informal teams.  After approximately
20~minutes of exploration, Q2 asked them to identify the peak-defect
time window (night, morning, afternoon, evening, or no clear
pattern).  After a further period of analysis, Q3 asked them to
classify the form of the defect signal (random noise, linear trend,
periodic/cyclic, step changes, or clusters).  These two polls
targeted progressively deeper analytical reasoning: Q2 required
temporal pattern recognition, while Q3 required structural
characterisation of the signal's shape.

\emph{Phase~3: Causal Integration} (Q4).  In the final poll, students
selected all factors they believed contributed to the defect increase
from six options (temperature control, humidity/moisture, material
contamination, equipment malfunction, operator shift patterns,
pressure variation).  This multiple-select format required students
to synthesise evidence from the preceding phases into a coherent
causal account.

Throughout the session, the instructor provided progressive hints at
timed intervals without revealing the root cause, and displayed
aggregated Mentimeter responses after each poll to enable whole-class
discussion.

\subsubsection{Data collection and analysis}

Mentimeter response data were exported for each of the four polls,
recording the number of respondents and the distribution of
selections.  Because students responded in informal teams, the unit
of analysis is the team-level response rather than the individual
student.  Response counts ($n = 52$, 32, 21, 9 across Q1--Q4)
therefore represent approximately one submission per team of three to
four students.  The declining counts across stages reflect both the
voluntary nature of the polls and the progressive consolidation of
within-team discussion, as teams that initially submitted multiple
individual responses increasingly converged on a single team
submission at later stages.

Within-session progression was assessed by tracking the percentage of
correct responses at each stage.  Two accuracy measures were computed: a strict measure (textbook-correct answer only) and an inclusive measure that additionally credited analytically defensible responses. The inclusive criterion was defined a priori from the dataset's process logic: because shift-linked humidity dynamics can manifest as step changes in aggregated defect data, ``step changes'' was judged a structurally valid characterisation even though ``periodic/cyclic'' is the conventional label (see Section~\ref{subsec:engagement}).  Fisher's exact test
was used to compare the proportion of correct responses between Q1
and Q4.  The student experience account
(Section~\ref{subsec:experience}) was solicited from the first author as a written reflection, lightly edited for clarity, and
analysed for convergence with the quantitative polling patterns.

\subsection{Outreach transferability study}\label{subsec:out_methods}

\subsubsection{Context and participants}

A condensed version of the framework was delivered as a four-hour
public engagement workshop for 26 Year~12 students (age 16--17) with
no prior engineering or data-analysis training.  The event was promoted to partner schools ranking highly on the university's engineering schools targeting list, which prioritises state-funded secondary schools and colleges within commuting distance where students are studying STEM A-levels. It was also promoted via the university's CRM system to prospective students who had previously engaged through fairs, open days, or online enquiries. The event was capped at 65 registrations, yielding 26 attendees (40\% conversion, slightly below the institutional average of 45--50\% for events of this type). Full details of participant demographics and school composition are reported in Section~\ref{subsubsec:context}.

\subsubsection{Adaptation}

The three-phase structure was preserved but the content was adapted
for a non-specialist audience (Table~\ref{tab:adaptation}).  Two
mystery scenarios replaced the polymer-specific dataset: Mystery~A
(phone screen cracking, humidity-driven---preserving the same root
cause as the undergraduate activity) and Mystery~B (football team
away performance, with travel distance as the hidden factor).  The
key design addition was a progressive AI-scaffolding sequence:
Mystery~A was completed without AI tools, followed by a Microsoft
Copilot demonstration, then Mystery~B with AI tools available.  This
enabled within-session comparison of manual versus AI-assisted
investigation.

\subsubsection{Instruments}

Five instruments were used:
(A)~matched pre/post Likert scales on three self-efficacy
items (Mentimeter, individual-level, $n = 10$ paired respondents);
(B)~Mystery~A case file (Microsoft Forms, team-level, $n = 9$);
(C)~Mystery~B case file (Microsoft Forms, team-level, $n = 10$);
(D)~post-event evaluation informed by the NERUPI (National Education Research and User Panel Initiative) framework for outreach effectiveness (Mentimeter, $n = 9$--$10$); and (E)~an
AI word cloud (Mentimeter, $n = 11$).  Pattern recognition accuracy
was assessed from the multiple-choice chart interpretation items.
Root-cause identification was coded from free-text team theories.
The cross-mystery difficulty comparison and AI attribution were
assessed from Mystery~B's additional items.

\subsubsection{Analysis}

Pre/post self-efficacy gains were tested using Wilcoxon signed-rank
tests on the $n = 10$ matched pairs, with effect sizes computed as
$r = |Z|/\sqrt{N}$.  Cross-mystery confidence was compared using the
same test on $n = 8$ matched teams.  Co-creation outputs were
analysed for concept coverage against a seven-item coding frame
(Table~\ref{tab:cocreation}) and classified against Bloom's revised
taxonomy.  All team-level free-text responses were coded by the corresponding author and checked against the raw
Microsoft Forms submissions.

\subsection{Ethical considerations}\label{subsec:ethics}

The undergraduate polling data were collected as part of routine
in-class formative assessment using anonymous Mentimeter responses;
no personally identifiable data were recorded.  The outreach workshop
was delivered as a public engagement activity under the university's
Centre for Public Engagement ethical guidelines.  Pre/post survey
responses were collected anonymously via Mentimeter, and team case
files were submitted voluntarily.  Participants were informed that anonymised, aggregated data might be used for educational research purposes. Outreach participants were minors (age 16--17) recruited through school partnerships; schools provided institutional consent for participation, and students attended voluntarily with parental awareness facilitated through the school registration process. No personally identifiable data were collected from outreach participants.

\section{Results}\label{sec:results}

This section presents evidence from two deployments of the detective
scaffolding framework.  Sections~\ref{subsec:progression}
and~\ref{subsec:engagement} report the quantitative polling data from
the primary undergraduate implementation;
Section~\ref{subsec:experience} complements these with a first-person
student account of the same session; and
Section~\ref{subsec:transferability} tests whether the framework transfers to a pre-university population with no prior engineering training. The core evidence derives from the undergraduate polling data; the student account and outreach study provide complementary qualitative and cross-context evidence respectively.

\subsection{Within-session learning progression}\label{subsec:progression}

\begin{figure}[!htbp]
\centering
\includegraphics[width=0.95\linewidth]{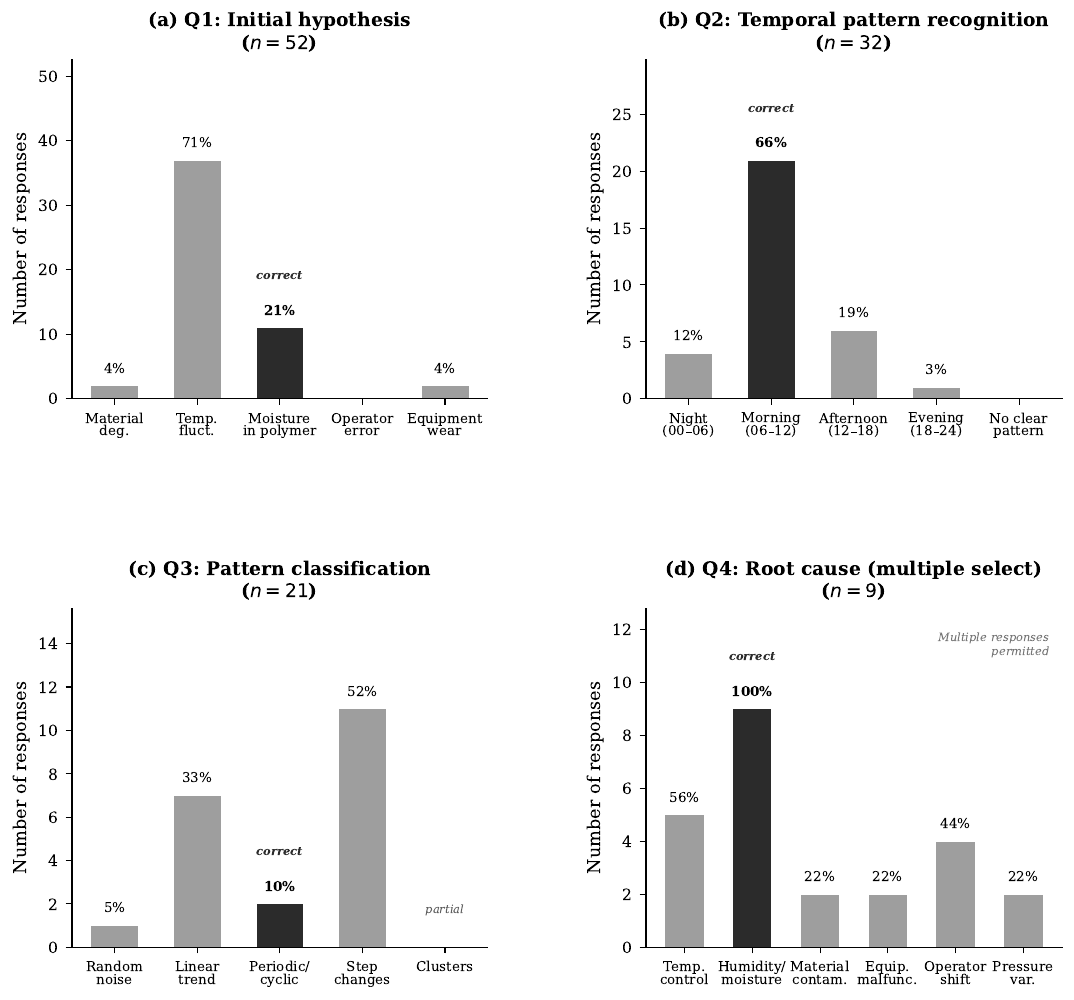}
\caption{Stage-wise response distributions across four investigative
prompts in the Polymer Detective activity.
\textbf{(a)}~Q1, initial hypothesis ($n=52$): 71\% attributed the
defect increase to temperature fluctuations; only 21\% selected
moisture (correct answer indicated).
\textbf{(b)}~Q2, temporal pattern recognition ($n=32$): 66\%
correctly identified the morning period (06:00--12:00) as the
peak-defect window.
\textbf{(c)}~Q3, pattern classification ($n=21$): responses shifted
toward structured descriptions---52\% selected step changes, 10\%
periodic/cyclic (correct); 33\% linear trend.
\textbf{(d)}~Q4, root-cause identification (multiple select, $n=9$):
100\% of respondents identified humidity/moisture as a contributing
factor.}
\label{fig:responses}
\end{figure}

Figure~\ref{fig:responses} presents the response distributions at each
of the four investigative stages.  A marked shift in causal
attribution is visible across the sequence.

At Q1 ($n=52$), temperature-based explanations dominated
(71\%), reflecting the strong prior association between temperature
and polymer processing defects established in preceding lectures.
Only 21\% of respondents identified moisture as a contributing factor
at this stage.  By Q4 ($n=9$, multiple-select), every remaining
respondent selected humidity or moisture, representing unanimous selection of humidity/moisture as a contributing factor among teams that completed the
full investigative sequence (Fisher's exact test, Q1 correct
vs.\ Q4 correct: $p < .001$; the odds ratio was undefined due to
complete convergence at Q4, with a Haldane-corrected estimate of
$\text{OR} \approx 69$). Because informal team composition may have shifted across stages and not all Q1 respondents also responded at Q4, this comparison reflects two cross-sectional snapshots rather than a within-subjects trajectory.

The intermediate stages reveal how this convergence was constructed
rather than guessed.  In Q2 ($n=32$), students examined time-resolved
defect data and 66\% correctly identified the 06:00--12:00 window as
the peak-defect period.  This question did not ask
students to explain the cause; it asked them to identify \emph{when}
defects concentrated.  The high correct-response rate suggests that
most teams had begun to engage in temporal reasoning---narrowing
the hypothesis space to variables that co-vary with time of day or
shift pattern---even before a causal mechanism was proposed.  This
marks a transition from static attribution (``temperature causes
defects'') to process-oriented analysis grounded in time-series
structure.

Q3 ($n=21$) required students to characterise the \emph{form} of the
defect signal rather than its timing.  Responses shifted toward
structured pattern descriptions: 52\% selected step changes and 10\%
selected periodic or cyclic behaviour (considered correct), while 33\%
described a linear trend.  The movement from ``when do defects peak?''
(Q2) to ``what shape does the signal take?'' (Q3) indicates a
qualitative shift from temporal localisation to structural abstraction.
Students selecting step changes or periodic patterns were no longer
searching for isolated anomalies but attempting to model the underlying
dynamics of the process---a necessary precursor to defensible causal
reasoning.

The complete convergence at Q4 therefore represents the outcome of
progressively constrained hypothesis refinement rather than a single
moment of insight.  The staged design---moving from open hypothesis
(Q1) through temporal evidence (Q2) and structural characterisation
(Q3) to causal synthesis (Q4)---mirrors the three-phase investigative
framework (Hypothesis Activation $\rightarrow$ Evidence Structuring
$\rightarrow$ Causal Integration) and provides structured polling
evidence that students' reasoning transformed qualitatively across
the session.

\FloatBarrier

\subsection{Engagement trajectory and participation}\label{subsec:engagement}

\begin{figure}[!htbp]
\centering
\includegraphics[width=0.85\linewidth]{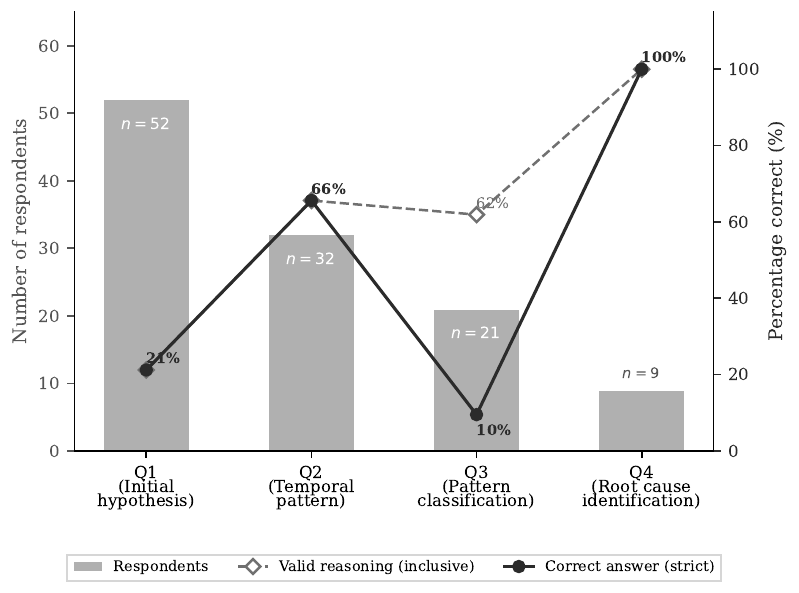}
\caption{Engagement and accuracy trajectory across the four
investigative stages.  Grey bars indicate the number of respondents at
each stage (left axis).  The solid line (\emph{correct answer, strict})
plots the percentage of respondents selecting the textbook-correct
option at each stage (right axis).  The dashed line (\emph{valid
reasoning, inclusive}) additionally credits structurally defensible
responses at Q3, where ``step changes'' (52\%) reflects valid
analytical reasoning about defect dynamics even though
``periodic/cyclic'' (10\%) was the conventional correct
classification.  The two lines diverge only at Q3 and converge at Q4,
where all remaining respondents identified humidity/moisture as the
root cause.}
\label{fig:engagement}
\end{figure}

While the preceding subsection focused on the \emph{content} of
responses, the pattern of \emph{participation} across stages also
warrants examination.
Figure~\ref{fig:engagement} summarises the relationship between
respondent numbers and accuracy across Q1--Q4.
Participation declined from 52 at Q1 to 9 at Q4, a pattern
consistent with the voluntary, non-assessed nature of the in-class
polls and the increasing analytical demand at each stage.

Two accuracy measures are plotted.  The strict measure records only the
textbook-correct response at each stage; the inclusive measure
additionally credits responses reflecting valid analytical reasoning.
The two measures diverge only at Q3, where the strict correct answer
(periodic/cyclic pattern, 10\%) was selected by few students, yet a
further 52\% identified step changes---a structurally defensible
characterisation of shift-related defect dynamics.  When step-change
responses are included, Q3 accuracy rises to 62\%, and the overall
trajectory shows a monotonic increase from 21\% to 100\%.

The divergence at Q3 is pedagogically informative rather than
problematic.  Students who selected ``step changes'' had moved beyond
temporal localisation (Q2) toward modelling the form of the defect
signal---the cognitive step the framework was designed to
elicit---even though their label did not match the textbook
classification.  This distinction matters for interpreting the learning
progression: the inclusive trajectory better represents the
\emph{quality of reasoning} at each stage, while the strict trajectory
captures convergence on canonical terminology.

The decline in participation warrants caution in interpreting Q4.
Complete convergence (100\% correct) was observed among the 9~teams
who remained engaged through the full sequence, but these
final-stage completers may represent a self-selected subset with
stronger data-analysis confidence or greater investment in the
investigative process.  Accordingly, the Q4 result should be
interpreted as evidence that the staged investigative design
\emph{supported} correct causal identification among persisting
teams, rather than as evidence that the entire cohort achieved
mastery.

\FloatBarrier

\subsection{Student experience}\label{subsec:experience}

The quantitative data presented above document \emph{what} students
concluded at each stage but cannot reveal \emph{how} the
investigative process was experienced from inside a working team.
This subsection offers that complementary perspective through a
first-person account by a student participant (first author).  The
account was written independently and has been lightly edited for
clarity; it is presented as an illustrative participant reflection offering qualitative evidence of the reasoning
processes that the polling data can only indirectly capture.

\subsubsection{Overall experiential impression}

The experience of participating in the Polymer Detective activity
differed substantially from most teaching activities I had previously
encountered in polymer courses.  Instead of being presented with a
well-defined problem and a prescribed analytical procedure, the
session began with an intentionally open-ended industrial scenario: a
sudden 15\% increase in defect rate in an injection-moulding facility,
accompanied by a dataset containing a large number of process and
environmental variables.  This deliberate ambiguity was initially
unsettling.  The volume and complexity of the data made it unclear
where to begin, and there was no obvious ``correct'' analytical
starting point.  However, with structured guidance built into the
activity, it became apparent that this initial uncertainty was not
accidental but central to the learning design.  Rather than being
guided toward a specific calculation or method, we were expected to
confront the problem as an authentic engineering task and to decide
for ourselves how it should be approached.

At the outset, my reasoning was strongly shaped by prior course
content.  Like the majority of the cohort, I initially attributed the
defects to temperature-related issues.  Within the context of polymer
processing, this assumption felt not only reasonable but almost
self-evident.  However, as we began systematically exploring the
dataset, this explanation became increasingly difficult to sustain.
Temperature variables appeared tightly controlled and showed little
meaningful correlation with defect rates.  This mismatch between
expectation and evidence produced a moment of discomfort: the data was
not confirming what I believed I understood.  In retrospect, this
tension between prior assumptions and empirical evidence defined my
overall experience of the activity: it marked a shift from
assumption-driven reasoning to evidence-based analysis grounded in
real production data.

\subsubsection{Group dynamics and collaborative reasoning}

Group work played a critical role at this stage.  Different members
pursued different analytical approaches---time-series visualisation,
correlation analysis, and grouping data by shift or time of day.
Discussion was not always efficient and occasionally felt fragmented,
but this divergence proved productive.  It prevented premature
convergence on a convenient explanation and forced each hypothesis to
be justified through evidence rather than intuition.  In hindsight,
what felt like inefficiency was essential to resisting an early but
incorrect conclusion.

\subsubsection{The moment of conceptual realisation}

The pivotal moment of the activity emerged gradually rather than as a
single dramatic reveal.  As patterns accumulated, we began to notice
that defect peaks aligned closely with shift changes, and that these
periods coincided with spikes in ambient humidity.  When defect rates,
time, and humidity were examined together, the pattern became
difficult to ignore.  The resulting realisation was powerful precisely
because it was self-generated: disparate observations finally cohered
into a plausible causal explanation.  This ``realisation moment'' was
not delivered by instruction, but earned through investigation.

Crucially, this insight felt tightly connected to the course content.
Concepts such as hygroscopic behaviour, plasticisation, and the
sensitivity of polymer melts to environmental conditions shifted from
abstract theory to practical explanatory tools.  Rather than recalling
definitions, we were applying polymer physics to justify why moisture
could plausibly drive defect formation in an industrial context.  This
alignment between theory and application made the learning experience
feel durable and meaningful.

\subsubsection{Critical reflection and perceived limitations}

The activity was not without challenges.  The open-ended nature of the
dataset was initially overwhelming, particularly during the first
15--20 minutes, and especially for students with limited confidence in
data analysis.  Additionally, variation in group progress created some
uneven pacing within the session.  A modest degree of early
scaffolding---such as suggesting initial exploratory analyses without
revealing the underlying cause---could help lower the entry barrier
while preserving the investigative character of the task.

Despite these limitations, the overall experience was highly engaging.
The detective framing reframed incorrect, incomplete hypotheses as a
natural part of the learning process rather than as failure.  This
acceptance of uncertainty closely mirrored real industrial
troubleshooting, where false leads and partial explanations are
expected.  As a result, the activity encouraged us to form a more
mature understanding of engineering problem-solving---one in which we
value rigorous reasoning under uncertainty more than arriving quickly
at the correct answer.

\medskip

Several elements of this account corroborate the quantitative findings
reported above.  The initial attribution to temperature aligns with the
71\% selection rate observed at Q1 (Fig.~\ref{fig:responses}a); the
description of gradual, evidence-driven convergence maps onto the
stage-wise progression from Q2 through Q4; and the report that group
divergence ``prevented premature convergence'' is consistent with the
intermediate heterogeneity visible in Q3 response distributions.  The
account also surfaces aspects the polling data cannot capture:
the affective experience of productive discomfort, the role of
within-group disagreement in sustaining investigation, and the
perceived connection between the activity and course content.  These
observations are revisited in the Discussion.

\FloatBarrier

\subsection{Cross-level transferability}\label{subsec:transferability}

The findings reported above demonstrate the framework's effectiveness
within its intended undergraduate context.  A natural follow-up
question is whether the pedagogical structure---rather than the
polymer-specific content---appears to be the primary driver of the observed learning progression.
To investigate this, a condensed version of the framework was deployed
with a population possessing no engineering background.

\subsubsection{Context and participants}\label{subsubsec:context}

Twenty-six Year~12 students (age 16--17) attended a four-hour
workshop at QMUL's Mile End Campus on 28 February 2026, recruited
through SEMS Student Recruitment and Marketing via school partnerships
from 65 registrants (40\% conversion).

School-level data were available for 14 of the 26 attendees across 11
schools.  The cohort was geographically dispersed across London and
South East England, including four comprehensive schools, five
selective or grammar schools, one independent school, and one
non-mainstream institution.  This demographic profile---academically
engaged, self-selected, and geographically diverse---is important for
interpreting the results. The cohort is not representative of a widening-participation population, despite QMUL having a strong widening-participation profile. However, this strengthens the transferability test. If students with no engineering background but
with generally strong academic profiles can successfully complete the
same investigation tasks designed for Year~3 undergraduates, the
framework's accessibility is demonstrated without confounding
questions about academic preparedness.

Students self-organised into 9--10 named teams.  One academic staff
member and three student ambassadors facilitated.

\subsubsection{Adaptation from undergraduate to outreach
context}\label{subsubsec:adaptation}

Table~\ref{tab:adaptation} summarises the key design changes.  The
workshop preserved the three-phase structure (Hypothesis Activation
$\rightarrow$ Evidence Structuring $\rightarrow$ Causal Integration)
but replaced polymer-specific content with accessible scenarios:
Mystery~A (phone screen cracking, humidity-driven) and Mystery~B
(football team away performance, travel distance as the hidden
factor).  The key pedagogical innovation was a progressive scaffolding
sequence: Mystery~A required manual chart interpretation, followed by
a Microsoft Copilot demonstration, then Mystery~B with AI tools
available, enabling within-session comparison of manual versus
AI-assisted investigation.

\begin{table}[!htbp]
\centering
\caption{Design adaptation from undergraduate module to outreach
workshop.}
\label{tab:adaptation}
\small
\begin{tabular}{@{}p{2.6cm}p{4.8cm}p{4.8cm}@{}}
\toprule
Element & Undergraduate (QXU5032) & Outreach \\
\midrule
Domain
  & Polymer engineering failure analysis
  & (A) Phone screen defects; (B) Football analytics \\
Data source
  & Purpose-built dataset (industrially informed)
  & Purpose-designed datasets with controlled variables \\
Duration
  & 90\,min (single lecture slot)
  & 4\,h (two mysteries + co-creation) \\
Scaffolding
  & Progressive Mentimeter polls (4 stages)
  & Mystery A manual $\rightarrow$ Copilot demo $\rightarrow$
    Mystery B AI-assisted \\
Prior knowledge
  & Year~3 polymer science
  & None \\
Assessment
  & Mentimeter polls (informal team submissions)
  & Team case file (Microsoft Forms) \\
Co-creation
  & None
  & ``Explain AI to a non-expert'' challenge \\
\bottomrule
\end{tabular}
\end{table}

\subsubsection{Pattern recognition and root-cause
identification}\label{subsubsec:patterns}

Table~\ref{tab:chartacc} reports chart-level interpretation accuracy.
In Mystery~A, all nine submitting teams (100\%) correctly identified
humidity as the primary cause, with 78\% identifying both contributing
factors (humidity and shift timing) and 67\% proposing a causal
mechanism (e.g., moisture weakening adhesive layers).  In Mystery~B,
all ten teams (100\%) identified travel distance, with every team
providing a multi-factor explanation integrating at least two
variables.  For the confounded variable (day of week), 60\% of teams
selected ``maybe, not sure''---arguably the most analytically
sophisticated response, as the day effect is confounded with travel
distance.  This suggests that even without formal training in
confounding variables, the structured investigation format prompted
appropriate epistemic caution.

\begin{table}[!htbp]
\centering
\caption{Chart interpretation accuracy by mystery.}
\label{tab:chartacc}
\small
\begin{tabular}{@{}llc@{}}
\toprule
Chart & Correct answer & Accuracy \\
\midrule
\multicolumn{3}{@{}l}{\emph{Mystery A --- Phone Screens ($n=9$ teams)}} \\
\quad Temperature vs Defects & No pattern & 9/9 (100\%) \\
\quad Humidity vs Defects & Clear pattern & 8/9 (89\%) \\
\quad Defects by Hour & Clear pattern & 6/9 (67\%) \\[4pt]
\multicolumn{3}{@{}l}{\emph{Mystery B --- Football Team ($n=10$ teams)}} \\
\quad Results by Day & Confounded & 6/10 ``Maybe'' (60\%) \\
\quad Home vs Away & Clear pattern & 9/10 (90\%) \\
\quad Travel vs Points & Clear pattern & 9/10 (90\%) \\
\bottomrule
\end{tabular}
\end{table}

These results mirror the undergraduate finding: unanimous identification of the key target factor among teams completing the full investigative sequence. That this was replicated with pre-university students
possessing no domain knowledge suggests that the progressive
scaffolding, rather than prior content expertise, is the primary
driver of the convergence effect.

\subsubsection{Pre/post self-efficacy gains}\label{subsubsec:prepost}

Ten participants completed matched pre- and post-activity Likert
items (Table~\ref{tab:prepost}).  AI explanation confidence and
data-analysis comfort both increased by more than one full Likert
point with large effect sizes ($r = .80$ and $.84$ respectively).  No
participant showed a decrease on any item.  Engineering interest
exhibited a ceiling effect (pre $M = 4.80$), consistent with the
self-selected cohort and therefore not interpretable as a null result.

\begin{table}[!htbp]
\centering
\caption{Paired pre/post self-efficacy results ($n = 10$).
Wilcoxon signed-rank tests; $r = |Z|/\sqrt{N}$.}
\label{tab:prepost}
\small
\begin{tabular}{@{}lcccccc@{}}
\toprule
Item & Pre $M$ (SD) & Post $M$ (SD) & $\Delta M$ & $W$ & $p$ & $r$ \\
\midrule
Explain AI to a friend
  & 3.20 (0.63) & 4.30 (0.82) & +1.10 & 0.0 & .008 & .80 \\
Interested in engineering
  & 4.80 (0.42) & 4.90 (0.32) & +0.10 & 0.0 & 1.00 & --- \\
Comfortable with data/patterns
  & 3.50 (0.53) & 4.60 (0.52) & +1.10 & 0.0 & .004 & .84 \\
\bottomrule
\end{tabular}
\end{table}

\subsubsection{AI-scaffolding effect}\label{subsubsec:aiscaffold}

Eighty percent of teams rated Mystery~B as easier than Mystery~A
(40\% ``much easier,'' 40\% ``a bit easier'').  Seven of ten teams
(70\%) spontaneously attributed this perceived ease to the AI tools
when asked to explain their reasoning, without being prompted to
evaluate the tools specifically.  Team confidence increased from
$M = 3.38$ (SD $= 1.06$) in Mystery~A to $M = 3.62$ (SD $= 1.06$)
in Mystery~B across eight matched teams, though this difference did
not reach significance (Wilcoxon $W = 2.5$, $p = .625$), likely due
to the small sample and ordinal-scale compression.

One team reported that it was ``harder to get the AI to work'' and was
the only team to rate Mystery~B as harder than Mystery~A, suggesting
that AI scaffolding is effective when functional but can become a
barrier when students encounter technical difficulties.

\subsubsection{Co-creation as evidence of higher-order
learning}\label{subsubsec:cocreation}

In a final activity, teams created explanations of AI for a general
audience.  Content analysis of four team outputs
(Table~\ref{tab:cocreation}) revealed that all four reached the
\emph{Create} level of Bloom's revised taxonomy, producing original
artefacts (an infographic, a website, a slide deck, and a hand-drawn
poster).  Two teams additionally demonstrated \emph{Evaluate}-level
reasoning: one built a complete educational website covering AI
classification, a six-step machine-learning pipeline, eight
application sectors, and a historical timeline---with no prior web
development experience; another produced a poster featuring a critical
analysis of AI's energy and water consumption impact on lower-income
countries.

\begin{table}[!htbp]
\centering
\caption{Concept coverage in co-creation outputs.}
\label{tab:cocreation}
\small
\begin{tabular}{@{}lccccc@{}}
\toprule
Concept & T1 & T2 & T3 & T4 & Total \\
\midrule
AI definition          & \checkmark & \checkmark & \checkmark & \checkmark & 4/4 \\
Applications/examples  & \checkmark & \checkmark & \checkmark & \checkmark & 4/4 \\
Limitations/risks      & \checkmark & \checkmark & \checkmark & \checkmark & 4/4 \\
How AI learns from data& \checkmark & \checkmark & \checkmark & ---        & 3/4 \\
Benefits vs risks      & \checkmark & \checkmark & ---        & \checkmark & 3/4 \\
Historical context     & ---        & \checkmark & ---        & ---        & 1/4 \\
Global equity critique & ---        & ---        & ---        & \checkmark & 1/4 \\
\bottomrule
\end{tabular}
\end{table}

\subsubsection{Post-event evaluation}\label{subsubsec:posteval}

Post-event Likert items ($n = 9$--$10$) confirmed high engagement:
``I see engineering as a path I could follow'' ($M = 4.90$),
``I would recommend this event'' ($M = 4.78$), and ``Engineering
feels interesting to study'' ($M = 4.70$).  Open-ended feedback
($n = 7$) identified the detective investigation activities as the
most valued element (5/7), with the most common improvement
suggestion being a request for more scenarios and more time (5/7).

\subsubsection{Methodological caveats}\label{subsubsec:outlimitations}

The paired analysis relies on $n = 10$ matched respondents from a
self-selected, academically engaged cohort (5 of 11 identified schools
were selective or grammar schools).  Response rates declined across
later survey items (pre 45\%, post 48\%, evaluation 27--30\%) due to
time pressure at the session close, which may introduce non-response
bias; however, participation in core activities remained high (82\% of
registered devices voted in the co-creation task, and 10 of $\sim$10
teams submitted Mystery~B case files).  The high baseline engineering
interest ($M = 4.80$) limits the ability to detect attitude change.
Team-level data ($n = 9$--$10$ teams) provides the more robust
evidence base.  The single-session design cannot address retention or
longer-term transfer.

\FloatBarrier

\section{Discussion}\label{sec:discussion}

The two deployments reported above produced converging evidence that
the detective scaffolding framework supports within-session reasoning
progression, but they differ in evidential strength.  The
undergraduate implementation provides the richer dataset: four
sequential polls capturing qualitative shifts in causal reasoning
among 80 students within a single session.  The outreach study
provides the transferability signal---identical root-cause
convergence with a non-specialist population---but relies on small
paired samples ($n = 10$) and team-level data ($n = 9$--$10$).  The
discussion that follows therefore draws primarily on the
undergraduate findings, using the outreach results to probe the
question of whether the pedagogical structure or the disciplinary
content is the active ingredient, while remaining cautious about the
generalisability of the outreach-specific claims.

\subsection{Cognitive engagement and productive failure}

The within-session polling data map onto the ICAP framework
\citep{Chi2014} at progressively higher engagement levels.  Q1
elicited an \emph{Active}-level response: students selected from
given options, drawing on prior knowledge without interrogating new
evidence.  Q2 and Q3 moved students into \emph{Constructive}
territory: they generated interpretations of temporal and structural
patterns that were not directly stated in the dataset.  The group
dynamics described in Section~\ref{subsec:experience}---where team
members pursued divergent analytical approaches and challenged each
other's hypotheses---reflect \emph{Interactive} engagement, the
highest ICAP level.  The framework's staged design was intended to
produce exactly this trajectory, and the polling data provide
empirical evidence that it did so.

The Q1 result---71\% of students attributing defects to
temperature---is better understood as pedagogically productive than as
a failure of prior instruction.  Within Kapur's productive failure
framework \citep{Kapur2008}, initial incorrect solutions that are
later revised through evidence produce deeper conceptual
understanding than correct solutions arrived at without struggle.  The
student account in Section~\ref{subsec:experience} describes this
dynamic explicitly: the ``moment of discomfort'' when temperature
explanations failed to fit the data preceded the gradual convergence
on humidity as the root cause.  The detective framing played a
specific role here: by casting incorrect hypotheses as investigative
leads rather than errors, it normalised the experience of being wrong
and sustained engagement through the revision process.

\subsection{Structured polls as formative evidence probes}

A methodological contribution of this work is the positioning of
in-class polls not merely as engagement tools but as purpose-built
evidence probes within an Evidence-Centred Design (ECD) logic
\citep{Mislevy2004}.  Each of the four Mentimeter prompts was
designed to elicit a specific reasoning capability: Q1 targeted
hypothesis generation from prior knowledge, Q2 targeted temporal
pattern recognition, Q3 targeted structural classification, and Q4
targeted causal synthesis.  This design aligns with the formative
assessment principle that effective feedback requires tasks calibrated
to make specific aspects of student thinking visible
\citep{Nicol2006, Shute2008}.

The dual-accuracy measure introduced at Q3 illustrates why this
framing matters.  Under conventional correct/incorrect scoring, Q3
accuracy was 10\%---apparently a regression from Q2's 66\%.  However,
the 52\% of students who selected ``step changes'' had demonstrably
moved beyond temporal localisation toward modelling the form of the
defect signal, the exact cognitive transition the framework intended
to elicit.  The inclusive measure (62\%) better represents the quality
of reasoning at that stage.  This distinction is important for
instructors adapting the framework: evaluating polls as evidence probes rather than quiz items changes what counts as a successful response and, consequently, what instructional decisions follow. This interpretation is consistent with principles of process-oriented assessment in engineering laboratory modules, where assessment focuses not on the correctness of final outcomes alone but on how students define problems, interpret data, and construct solutions. From this perspective, a response such as ``step changes'' at Q3---while not the canonical label---constitutes visible evidence of an intermediate reasoning stage that correctness-based scoring would obscure.

\subsection{Positioning against prior work}

The detective framing builds on an established tradition of
gamified investigation in science education, including escape rooms
in engineering \citep{Veldkamp2020} and forensic case studies in
chemistry \citep{Dinan2007}.  More recently, puzzle-based
polymer learning environments have demonstrated that investigative
formats can be adapted to materials science contexts
\citep{Gilbert2020}.  The present study extends this work in three
directions.  First, it operates at large-cohort scale (80~students
in a single lecture slot), whereas most escape-room implementations
involve 20--30 participants in dedicated studio settings.  Second, it
uses industrially realistic datasets with process variables modelled
on manufacturing experience
rather than purpose-designed puzzles, aligning with arguments for
``messy'' data in science education \citep{Kjelvik2019}.  Third,
and most distinctively, it provides within-session progression
evidence through staged polling---most comparable studies report only
endpoint outcomes such as final-exam scores or satisfaction surveys
\citep{Freeman2014}.

The large-cohort constraint is particularly relevant for the
discipline.  Problem-based learning frameworks are well-developed
\citep{Savery2006}, but implementation typically assumes smaller
cohorts and longer contact time.  The detective framework demonstrates
that structured investigation can fit within a standard 90-minute
lecture slot for 80+~students, using Mentimeter polls and informal
team structures to maintain feasibility without dedicated PBL
infrastructure.

\subsection{Transferability and the scaffolding--content distinction}

The outreach study's most important implication concerns what drives
the observed learning effect.  The Year~12 participants possessed no
polymer science knowledge, yet they achieved identical root-cause
identification rates to the undergraduates across two distinct
scenarios.  This suggests that the three-phase structure itself---the
progression from open hypothesis through systematic evidence
evaluation to causal synthesis---is sufficient to produce convergent
reasoning, independent of prior disciplinary expertise.  The
practical consequence is that the framework is portable: instructors
in other disciplines could adapt the detective scaffolding to their
own datasets and contexts while preserving the pedagogical mechanism.
The investigation also mirrors the diagnostic reasoning required in
industrial quality assurance, suggesting that the framework develops
transferable professional skills alongside disciplinary understanding.

The outreach deployment also surfaced an emergent finding.  The
manual-first design (Mystery~A without AI tools, Mystery~B with
Microsoft Copilot) ensured that students first experienced the
cognitive demands of data investigation before receiving AI
scaffolding.  Seventy percent of teams spontaneously credited AI
tools with improved performance, without being prompted to evaluate
them.  This unprompted attribution suggests that students could
articulate the specific value AI added---pattern identification,
efficiency, reduced cognitive load---precisely because they had a
concrete baseline experience for comparison.  The detective framework
may therefore serve as a vehicle for developing AI literacy alongside
disciplinary reasoning, a possibility that warrants further
investigation as AI-assisted pedagogy becomes more prevalent.

The co-creation outputs provide additional evidence of the framework's
capacity to support higher-order learning.  That four teams
independently produced original artefacts covering AI definitions,
applications, \emph{and} limitations---with two teams reaching
evaluative critique of AI's societal impact---demonstrates
progression beyond investigation to knowledge synthesis within a
single compressed session.

\subsection{Limitations and future directions}

Four limitations should be acknowledged, each pointing to a specific
extension of the present work.

First, participation declined from 52~team-level responses at Q1 to 9
at Q4 in the undergraduate implementation.  The team-based response
structure (3--4 students per submission) means these figures understate
individual engagement, but the Q4 convergence nonetheless reflects
persisting teams rather than the full cohort.  Whether non-responding
teams also converged on the correct root cause cannot be determined
from the available data.  Tagging each poll response by knowledge
component would enable Bayesian Knowledge Tracing (BKT) to model
skill-level mastery trajectories across cohorts \citep{Corbett1995,
SaricGrgic2024}.  The current four-prompt design provides an initial
progression signal; a finer-grained item structure would allow
instructors to identify which specific reasoning transitions (e.g.,
temporal to structural) individual students find most difficult,
enabling targeted within-session intervention---and would provide a
richer basis for characterising the reasoning of teams that disengage
before Q4.

Second, the outreach transferability study relies on $n = 10$ paired
respondents from a self-selected, academically engaged cohort.  The
large effect sizes ($r > .79$) provide confidence in the direction
and magnitude of the gains, but the sample limits generalisability
to less engaged or less academically prepared populations.  A planned
extension addresses this directly: the Spectral Detective activity---a
pilot version applied to Raman spectral classification---was tested
with too few respondents ($n \leq 4$) to draw conclusions in this
study.  A full-scale deployment with adequate sample sizes would test
whether the detective scaffolding produces equivalent reasoning shifts
in instrument-based analytical tasks, where the evidence is
spectroscopic data rather than process variables, and would broaden
the evidence base beyond the populations reported here.

Third, both deployments assessed reasoning within a single session.
Within-session progression is not equivalent to durable learning.
Whether students who participated in the detective activity demonstrate
different performance in subsequent modules, or retain the causal
reasoning skills developed during the session, remains an open
empirical question.  Longitudinal tracking---comparing performance in
subsequent laboratory modules or final examinations with matched peers
who did not participate---would address this gap and strengthen the
case for the framework's lasting pedagogical value.

Fourth, the study reports a single-site, single-module implementation.
Replication across institutions, disciplines, and cultural contexts
would be needed to establish the framework's generalisability beyond
the specific conditions reported here.  The cross-level transferability
result provides an initial signal that the pedagogical structure is
portable, but confirmatory deployments by independent instructors in
different disciplinary and institutional settings remain a priority.

\section{Conclusion}\label{sec:conclusion}

This study presented and evaluated a three-phase detective
scaffolding framework for polymer engineering education in which
staged in-class polls function as designed evidence probes rather
than engagement tools.  Among 80 undergraduate students, the
framework produced a within-session progression from
prior-knowledge-driven misconception (71\% attributing defects to
temperature) to unanimous selection of humidity/moisture as a key contributing factor among final responding teams, with intermediate stages revealing qualitative
shifts in reasoning from temporal localisation to structural
abstraction.  Among 26 pre-university students with no engineering
background, the same three-phase structure produced identical
root-cause identification rates across two distinct scenarios, with
significant gains in data-analysis confidence and AI explanation
ability.

Three implications follow for science and engineering education
practice.  First, the detective framing demonstrates that narrative
investigation and large-cohort lecture delivery are not
incompatible: the framework operated within a standard 90-minute slot
for 80~students using only Mentimeter and informal team structures,
without requiring dedicated PBL infrastructure.  Second, the
dual-accuracy finding at Q3---where textbook-correct and
analytically valid responses diverged---illustrates a broader
principle: evaluating formative polls as evidence probes rather than
quiz items changes what counts as a successful response and,
consequently, what instructional decisions follow.  Third, the
cross-level transferability result suggests that the pedagogical structure, rather than the disciplinary content, appears to contribute importantly to the convergence effect, indicating that the framework may be adaptable to other disciplines and
educational levels provided the core three-phase sequence is
preserved.

Future work should extend this study in three directions: integrating
Bayesian Knowledge Tracing to model skill-level mastery trajectories
from poll response data, deploying the framework at full scale in
instrument-based analytical contexts (e.g., spectral classification),
and tracking whether within-session reasoning gains persist into
subsequent modules.

\section*{Availability of data and materials}
Aggregated, anonymised response data supporting this study are openly
available from Figshare at \url{https://doi.org/10.6084/m9.figshare.31921983}.

\section*{Ethics declaration}
The undergraduate polling data were collected as part of routine in-class
formative assessment using anonymous responses; no personally identifiable
data were recorded. The outreach workshop was delivered as a public
engagement activity under the university's Centre for Public
Engagement ethical guidelines. Outreach participants were minors (age
16--17) recruited through school partnerships; schools provided
institutional consent for participation, and students attended voluntarily
with parental awareness facilitated through the school registration
process. No personally identifiable data were collected from outreach
participants.

\section*{Disclosure statement}
The authors report there are no competing interests to declare.

\section*{Funding}
This work was supported by a Centre for Public Engagement Small Grant,
Queen Mary University of London (2025--26).

\section*{Authors' contributions (CRediT)}
\textbf{Haolin Feng}: Investigation, Writing -- original draft
(student experience section), Visualization, Writing -- review \& editing.
\textbf{Holly Barrett}: Resources (school recruitment and outreach
logistics), Project administration (workshop delivery).
\textbf{Xinru Deng}: Conceptualization (assessment design discussions
informing the formative evidence structure).
\textbf{Dimitrios G.\ Papageorgiou}: Supervision (co-taught the
Physical Properties of Polymers module), Writing -- review \& editing.
\textbf{Yiwei Sun}: Conceptualization, Methodology, Formal analysis,
Investigation, Data curation, Writing -- original draft, Writing --
review \& editing, Supervision, Project administration, Funding
acquisition.

\section*{Biographical notes}

\textbf{Haolin Feng} is a Year~3 undergraduate student in Materials
Science and Engineering at Queen Mary University of London (QMUL),
studying on the joint UK--China programme with Northwestern
Polytechnical University.  His research interests include polymer
processing, active learning, and engineering education.

\textbf{Holly Barrett} works in Student Recruitment and Marketing in
the School of Engineering and Materials Science at QMUL, where she
leads school partnerships and coordinates outreach events aimed at
widening participation in engineering.

\textbf{Xinru Deng} is a Lecturer in Materials Science Education in
the School of Engineering and Materials Science at QMUL.  Her work
focuses on assessment design, team-based learning, and curriculum
development for joint UK--China engineering programmes.

\textbf{Dimitrios G.\ Papageorgiou} is a Reader in
Materials Science in the School of Engineering and Materials Science at
QMUL.  His research interests span polymer nanocomposites, mechanical
properties of polymers, and engineering education.

\textbf{Yiwei Sun} is a Lecturer in Materials Science and Engineering
Education in the School of Engineering and Materials Science at QMUL.
His scholarship focuses on learning analytics, algorithmic team
formation, and active learning design in large-cohort engineering
modules.  He is the corresponding author of this work.

\bibliographystyle{apalike}
\bibliography{references_in_order}

\end{document}